\documentclass{PoS}
\usepackage{fixmath,amsmath,amssymb}

\newcommand{\VEC}[1]{\mathbf{#1}}

\title{Moving NRQCD and $\mathbold{B \to K^* \gamma}$}

\ShortTitle{Moving NRQCD and $B \to K^*\gamma$}

\author{Stefan~Meinel, Ronald~Horgan, Lew~Khomskii,
  Laurent~C.~Storoni, and \speaker{Matthew~Wingate}\\
  DAMTP, University of Cambridge, Wilberforce Road, Cambridge CB3 0WA, UK\\
  E-mail: \email{M.Wingate@damtp.cam.ac.uk}}

\abstract{The formulation of NRQCD discretized in a reference frame boosted
  relative to the $B$ rest frame will enable calculation of $B$ form factors
  over a larger range of momentum transfer.  We have initiated a program to
  calculate form factors describing the rare decay $B \to K^* \gamma$. We
  discuss the strategy and challenges of the project.  As a first step in the
  numerical calculations, we present first results for bottomonium quantities
  using the $O(\Lambda_{\scriptscriptstyle\mathrm{QCD}}^2/m^2,
  v_{\scriptscriptstyle\mathrm{NR}}^4)$ moving NRQCD action.}

\FullConference{The XXV International Symposium on Lattice Field Theory\\
		 30 July-4 August 2007\\
		 Regensburg, Germany}

\begin{document}
\bibliographystyle{apsrev}

\section{Motivation}
\label{sec:motivate}

We at this conference know very well the importance lattice QCD calculations
have in the global flavor physics program.  Calculations of the $B$ meson
decay constant, $B\to\pi$ form factors, and $B-\overline{B}$ mixing 
matrix elements have been pursued and refined for over a decade,
and they are important ingredients in constraining 
parameters governing quark flavor-changing interactions.

It is now clear that the CKM mechanism of the Standard Model accurately
describes flavor physics up to present precision.  In order to probe
the couplings to the non-Standard Model physics we expect, we must
further refine experimental measurements and theoretical calculations.

In the latter pursuit, lattice QCD must extend its focus.  Rare $B$ decays
offer a promising avenue for improvement from the status quo.  One difference
between the rare $B$ decays and the processes on which lattice QCD usually
focuses is that the former require more assumptions, \textit{e.g.}\ neglect of
long distance contributions and hard spectator effects.  Nevertheless, lattice
calculations can still play an important role in the phenomenology of
exclusive $b\to s$ decays by reducing uncertainties in hadronic
matrix elements.

\section{Plan for calculation}
\label{sec:plan}

\begin{table}
\begin{center}
\begin{tabular}{ccl}
Matrix element & Form factor & Relevant decay(s) \\ \hline \\[-4mm]
$\langle P|\bar{q}\gamma^\mu b|B\rangle$ & $f_+, f_0$ & 
$\left\{\begin{array}{l} B\to\pi\ell\nu\\B\to K\ell^+\ell^-\end{array}\right.$
\\[4mm]
$\langle P|\bar{q}\sigma^{\mu\nu}q_\nu b|B\rangle$ & $f_T$ &
\hspace{3.5mm}$B\to K\ell^+\ell^-$ \\[4mm]
$\begin{array}{c}\langle V|\bar{q}\gamma^\mu b|B\rangle
\\ \langle V|\bar{q}\gamma^\mu\gamma^5 b|B\rangle\end{array}$ & 
$\begin{array}{c}V\\ A_0, A_1, A_2\end{array}$ &
$\left\{\begin{array}{l} B\to(\rho/\omega)\ell\nu \\
B\to K^*\ell^+\ell^-\end{array}\right.$ \\[7mm]
$\begin{array}{c}\langle V|\bar{q}\sigma^{\mu\nu}q_\nu b|B\rangle \\
\langle V|\bar{q}\sigma^{\mu\nu}\gamma^5 q_\nu b|B\rangle\end{array}$ &
$\begin{array}{c}T_1\\T_2, T_3\end{array}$ &
$\left\{\begin{array}{l} B\to K^*\gamma \\
B\to K^*\ell^+\ell^-\end{array}\right.$
\end{tabular}
\end{center}
\caption{Full list of $B$ semileptonic form factors.}
\label{tab:formfact}
\end{table}

In this section we outline our strategy for computing $B\to K^* \gamma$ form
factors.  Ultimately we would like to calculate all of the semileptonic $B$
decay form factors (Table~\ref{tab:formfact}).  Presently we concentrate on
the radiative decay because it stands to be the most greatly improved.

The main new component to be used is moving NRQCD (mNRQCD).  As with
conventional NRQCD, this is an effective field theory which permits lattice
calculations with the physical bottom quark mass.  The formulation in a frame
where the lattice is boosted relative to the $B$ rest frame will permit
calculations over a larger range of momentum transfer $q^2$ than non-moving
NRQCD.  We discuss mNRQCD in Section~\ref{sec:mnrqcd}.

We will use an improved staggered quark action for the light valence and
sea quarks.  The first calculations will make use of the ensemble of MILC
configurations generated with the AsqTad action; later we will use 
configurations generated with the HISQ action.  The virtues and risks
of using rooted, improved staggered quarks have been discussed extensively
\cite{Sharpe:2006re,Kronfeld:latt2007}.  A few remarks regarding the
$K^*$ are made in Section~\ref{sec:vecmeson}.

The matching between the continuum and lattice current and penguin operators 
will be carried out to 1-loop order in perturbation theory.  The matching
of the vector and axial vector currents for mNRQCD is being finalized presently
\cite{Khomskii:inprog}, and the matching for the penguin operator is underway.

A recent lattice calculation used a very different lattice strategy to
calculate the $B\to K^*$ form factors \cite{Becirevic:2006nm} (see within for
earlier lattice calculations).  The use of many approaches, sum rules in
addition to lattice QCD, is especially desirable given the theoretical
uncertainties.

\section{Moving NRQCD}
\label{sec:mnrqcd}

Moving HQET/NRQCD has been a recurring topic for over a decade
\cite{Hashimoto:1995in,Sloan:1997fc,Foley:2002qv,Boyle:2003ui,Dougall:2005zh,Davies:2006aa}.
Initially it was envisioned for use calculating Isgur-Wise functions at
nonzero recoil.  Since the $B\to D$ form factor shapes are constrained by
dispersion relations accurately, only the zero recoil normalization is now
necessary from lattice QCD (LQCD).  Later, mNRQCD was explored with the idea
of extending the reach of LQCD calculations of $B\to \pi$ form factors toward
large recoil.  This is still desirable, but the shape is now being measured
competitively by experiment.  In the previous 2 cases the LQCD determination
of the shape is not imperative, but the LQCD determination of the
normalization is still needed.  On the contrary, in order to reach the
physical point for $B\to K^*\gamma$ ($q^2=0$) where LQCD can provide the
normalization, a lattice calculation of the shape is a necessary step.  Moving
NRQCD is an important tool to develop and apply.

As with NRQCD, we work with an effective field theory which requires $m_b >
1/a$.  This condition is satisfied on all present and near-future unquenched
lattices.  Although one cannot take a continuum limit in the formal sense, we
can study and remove discretization errors at least as well as with other
heavy quark formulations.  There is no \textit{theoretical} problem with
working with a finite lattice spacing either. There are no discretization
errors on the renormalized trajectory.  Of course one can question how close
to the renormalized trajectory we can get using the Symanzik improvement
program.  However, this is a \textit{practical} question, the type of which
can be asked of any lattice formulation and can only be answered empirically.
Experience has shown NRQCD to be a successful approach.

The lattice (m)NRQCD action can be used for both
$\Upsilon$ and $B$ physics.  In the latter case, we use standard HQET
power counting to order and match operators.  The leading uncertainty
in some cases is the matching, done with 1-loop perturbation theory so far.
The convergence of HQET worsens as the recoil momentum becomes much 
larger than $\Lambda_{\scriptscriptstyle\mathrm{QCD}}$; however, we expect
the change to be mild over the range of $q^2 > 0$ we plan to study directly.

Working with a lattice boosted with respect to the $B$ meson has the potential
to blur the separation between physical and lattice length scales.  At rest,
hadronic momenta are of order $\Lambda_{\scriptscriptstyle\mathrm{QCD}}$.  In
a frame where the $B$ is boosted with velocity $v$, the boosted momentum
distribution is of order $\Lambda_{\scriptscriptstyle\mathrm{QCD}}
\sqrt{(1+v)/(1-v)}$ in the direction parallel to $\VEC{v}$.  That is,
discretization errors will be twice as large at $v = 0.6$ than for non-moving
NRQCD.  We anticipate that other sources of error will still dominate.

We have independently derived, coded, and tested the moving NRQCD action
accurate through $O(\Lambda_{\scriptscriptstyle\mathrm{QCD}}^2/m^2)$ for $B$
physics (HQET counting) and $O(v_{\scriptscriptstyle\mathrm{NR}}^4)$ (NRQCD
counting) for $\Upsilon$ physics.  The primary goal of our $\Upsilon$
calculations with mNRQCD is to test the code, checking that we obtain sensible
results with reasonable statistical errors as the boost velocity $v$
increases.  As far as we are aware, these are the first mNRQCD calculations
with a Lagrangian of this accuracy.  These tests were performed on a subset of
$2+1$ flavor AsqTad-fermion lattices provided by the MILC Collaboration, with
$\beta = 6.76$, bare quark masses 0.007 and 0.05, $V= 20^3\times 64$
\cite{Aubin:2004wf}.  We used the bare heavy quark mass, $a m=2.8$, which gave
the correct $B_s$ and $\Upsilon$ kinetic masses using non-moving NRQCD
\cite{Wingate:2003gm,Gray:2005ur}.

\begin{figure}
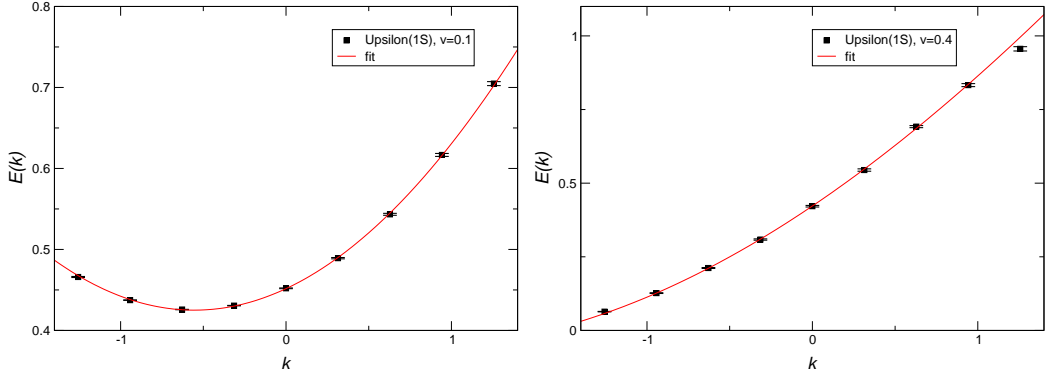
 
\begin{center}
\includegraphics[width=.45\textwidth]{disp_rel_v_01.eps} 
~\includegraphics[width=.45\textwidth]{disp_rel_v_04.eps} 
\caption{Preliminary dispersion relation $E(k)$ as a function of residual
momentum $k$, both in lattice units. The bare boost velocity is
$v=0.1$ (left) and $v=0.4$ (right).}
\label{fig:dispersion} 
\end{center}
\end{figure}

First we studied how spectral quantities behaved as the boost velocity $v$
varied.  On Coulomb gauge-fixed lattices, we used smeared interpolating
operators of the form
\begin{equation}
  O_v(\VEC{x},\tau) = \sum_{\VEC{r}} \overline{\Psi}_v
  (\VEC{x},\tau)\: f(\VEC{r})\Gamma\: \Psi_v(\VEC{x}+\VEC{r},
  \tau), 
  \label{eqn:bot_intpl_fields}
\end{equation}
where $f(\VEC{r})$ is a radial smearing function and $\Gamma$ is a
Dirac $\gamma$ matrix.  As in non-moving NRQCD, we decouple the
quark and antiquark fields $\Psi_v = (\psi_v, \chi_v)^T$ and evolve
the propagators from the source timeslice to the sink timeslice.
At the sink we project onto residual meson momentum $\VEC{k}$.
The energies can then be fit to
\begin{equation}
  E(\VEC{k}) = \sqrt{(2\gamma m \VEC{v} Z_p  +\VEC{k})^2 
    + M_\mathrm{kin}^2} + \Delta_v
\label{eq:dispersion}
\end{equation}
where $M_\mathrm{kin}$ is the kinetic meson mass, and $\Delta_v$ is
an additive energy shift which is a function of $v$ and is the same
for all mesons.  Note the physical meson momentum is split into
a residual momentum $\VEC{k}$, present explicitly in the calculation of
the correlation function, and an external momentum $2\gamma\,m \VEC{v}Z_p$,
with $\gamma = (1-v^2)^{-1/2}$.  $Z_p$ accounts for renormalization of the 
external momentum; we always find it to be consistent with 1 within
fitting uncertainties.
Dispersion relations for $\eta_b(1\mathrm{S})$ and $\Upsilon(1\mathrm{S})$
for different boost velocities are plotted in Figure~\ref{fig:dispersion}.

\begin{figure} 
\begin{center}
\includegraphics[width=.6\textwidth]{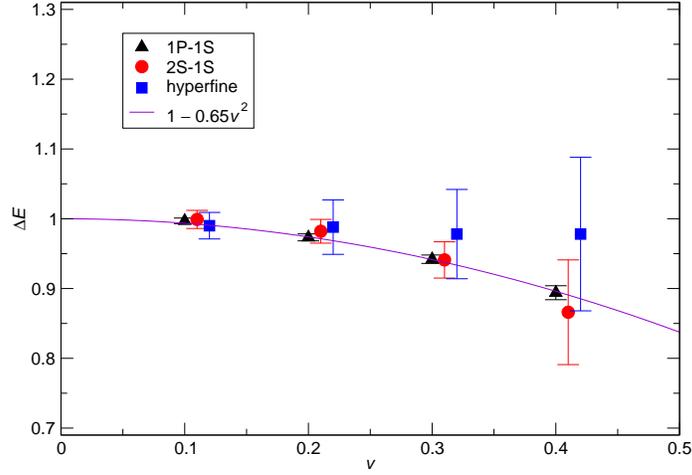} 
\caption{Preliminary bottomonium energy splittings $\Delta E$ as a function of
  boost velocity $v$, plotted relative to $\Delta E$ computed with $v=0$.
  (Points are offset horizontally for legibility.)  The 1P$-$1S and 2S$-$1S
  splittings show a $1 - c v^2$ decrease as expected from the dispersion
  relation.}
\label{fig:splittings} 
\end{center}
\end{figure}

In Figure~\ref{fig:splittings} we show several energy splittings as a
function of $v$, computed using correlation functions which project onto
residual momentum $\VEC{k} = 0$.   We note the statistical errors grow 
as $v$ increases from $0$ to $0.4$, an effect more pronounced for the
hyperfine and 2S$-$1S splittings than the 1P$-$1S splitting.
Splittings with non-moving NRQCD were computed in \cite{Gray:2005ur}.

Finally, to go beyond energies to matrix elements, we computed the $\eta_b$ to
vacuum matrix element of a fictitious axial vector current $A^\mu(x) =
\overline{\Psi}(x)\gamma_5\gamma^\mu\Psi(x)$, which we parametrize with a
decay constant $f$ as
\begin{equation}
\langle 0 | {A}^\mu(0) |\eta_b(1S),\VEC{p}\rangle = if\:p^\mu
\end{equation}
(in Minkowski spacetime).  The appropriate correlation function is constructed
by writing this operator in terms of the mNRQCD fields (in the lattice rest
frame) using the following transformation:
\begin{equation}
\Psi(x)\:=\:S_\Lambda\:T_{\scriptscriptstyle
  \mathrm{FWT}}\:\:\:e^{-im\:u\cdot x\:\gamma^0} \:
  T_{\scriptscriptstyle\mathrm{TD}}\:\:\frac{1}{\sqrt{\gamma}}\: \Psi_v(x)
\label{eqn:MNRQCD_field_redef}
\end{equation}
where
\begin{equation}
T_{\scriptscriptstyle\mathrm{FWT}}=\exp\left(\frac{i}{2m}\:\gamma^j 
\Lambda^\mu_{\:\:\:j}D_\mu\right)...
\end{equation}
is the Foldy-Wouthuysen-Tani transformation in the boosted frame,
\begin{equation}
T_{\scriptscriptstyle\mathrm{TD}}=\exp\left(\frac{i}{4\gamma m}\gamma^0
\left[(\gamma^2-1) D_0+(\gamma^2+1)\VEC{v}\cdot\VEC{D}\right]\right)...
\end{equation}
removes unwanted time derivatives, and $S_\Lambda$ is the Dirac spinor
representation of the Lorentz boost.

\begin{figure} 
\begin{center}
\includegraphics[width=.6\textwidth]{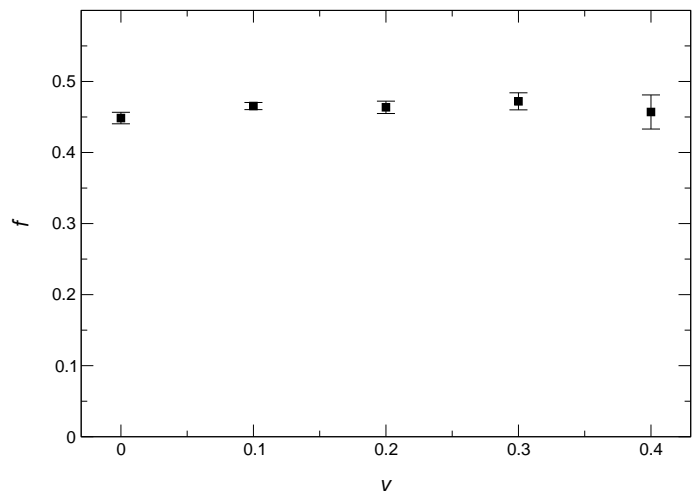} 
\caption{Preliminary results for the $\eta_b$ decay constant $f$, in lattice
  units, as a function of bare boost velocity.}
\label{fig:decayconst} 
\end{center}
\end{figure}

Figure~\ref{fig:decayconst} shows this decay constant computed for several
boost velocities.  We might expect some dependence on $v$ due to 
$v$-dependent operator renormalization and the fact that constant
bare heavy quark mass might not correspond to constant $M_{\eta_b}$.
Nevertheless, $f$ appears independent of $v$ within the statistical errors.

We note the statistical error increases by a factor of 3.  Increasing the
signal-to-noise ratio for correlators computed with $v>0$ will be an important
challenge for our planned matrix element calculations.  Progress has already
been achieved for $B\to \pi$ form factors (in the $v=0$ frame) by using random
wall sources \cite{Davies:2007vb}.

\section{Vector meson final state}
\label{sec:vecmeson}

\begin{figure} 
\begin{center}
\includegraphics[width=.6\textwidth]{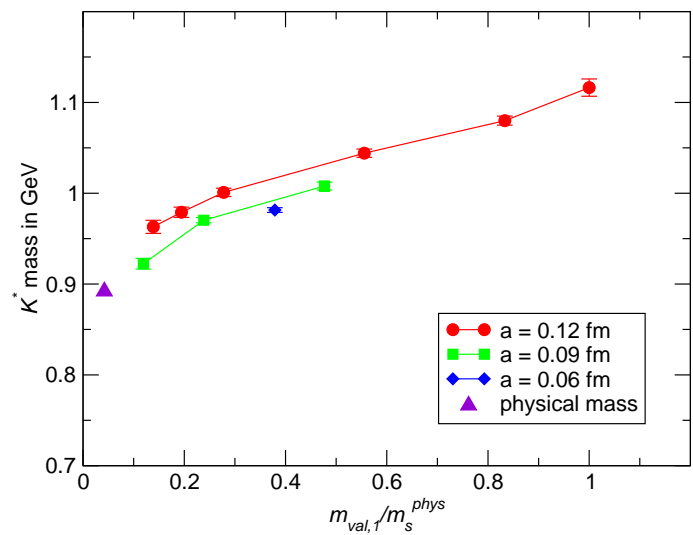} 
\caption{Unquenched $K^*$ mass as a function of light quark mass on MILC
  lattices (3 spacings), after simple interpolation to physical strange quark
  mass \cite{Aubin:2004wf}.  (Raw data communicated by D.~Toussaint.)
  Although statistically significant, scaling violations are small compared
  to other errors anticipated for $B\to K^*$ form factors.}
\label{fig:mkstarinterp} 
\end{center}
\end{figure} 

Figure~\ref{fig:mkstarinterp} shows the $K^*$ mass computed by the MILC
Collaboration \cite{Aubin:2004wf}.  Discretization errors are visible within
the small statistical errors, but are only a few percent, much smaller than
the other systematic errors we anticipate for the form factors.  Taste
splitting effects are negligible between the vector meson masses computed
with local and 1-link operators.

There are interesting issues to study regarding threshold effects as the quark
mass decreases.  Our initial calculations will be done with parameters for
which the $K^*$ is a stable state.  (Note that experimentalists quote
branching ratios which treat the vector resonance as a final state.)  Given
that we do not have a low energy effective theory for the vector mesons, as we
do for the pseudoscalar mesons and baryons, the best we can do is empirically
extrapolate from our input quark masses to the physical point.  The $B\to\pi$
form factors have a very mild quark mass dependence, so it is reasonable to
expect the same of the $B\to K^*$ form factors, up to threshold effects.  

\section{Conclusions}
\label{sec:concl}

Although more complicated than the standard $B$ meson matrix elements
calculated on the lattice, matrix elements relevant for rare $B$ meson decays
are increasingly important to the flavor physics program.  The difficulties
involved call for investigation with new tools such as moving NRQCD.  We have
implemented and tested the mNRQCD action through
$O(\Lambda_{\scriptscriptstyle\mathrm{QCD}}^2/m^2,
v_{\scriptscriptstyle\mathrm{NR}}^4)$.  We present here preliminary results
with this action, concentrating on the bottomonium dispersion relation, level
splittings, and the $\eta_b$ decay constant.  We are now working on
calculations for $B$ mesons.

\section*{Acknowledgments}

We thank P.~Ball, C.~T.~H.~Davies, E.~Gardi, A.~Hart, E.~M\"uller,
J.~Shigemitsu, K.~Wong, and R.~Zwicky for discussions.  This work was
supported by STFC.

\bibliography{mbw}

\end{document}